\documentstyle[twocolumn,prb,aps]{revtex}
\begin{document}
\bibliographystyle{prsty}
\draft
\title
{Ground State of a Spin System with Two- and Four-spin Exchange Interactions
 on the Triangular Lattice }

\author
{Kenn  Kubo and Tsutomu Momoi}

\address
{Institute of Physics, University of Tsukuba, Tsukuba, Ibaraki 305, Japan}

\date
{\today}

\maketitle
\begin{abstract}
We study  a spin system with both two- and four-spin
exchange interactions on the triangular lattice as a possible model for the 
nuclear magnetism of solid $^3$He layers adsorbed on grafoil. 
The ground state is analyzed by the use of the mean-field approximation. 
It is shown that the four-sublattice 
state is favored by introduction of the four-spin exchange interaction. 
A possible phase transition 
at a finite temperature into a phase with the scalar chirality is predicted.
Application of a magnetic field is shown to cause a variety of phase 
transitions.
\end{abstract}

\vspace{1cm}

It is a great pleasure to dedicate this paper to Professor Wolfgang G\"otze
for his sixtieth birthday, in whose research group at Max Planck Institut 
f\"ur Physik und Astrophysik and Die Technische Universit\"at M\"unchen 
one of the authors (K.K.) had the privilege to spend very pleasant and 
profitable years as a postdoctoral research associate, and to wish him 
many more happy years of activity in theoretical physics.
\section{Introduction}
Nuclear magnetism of the two-dimensional $^3$He layers adsorbed on 
a grafoil has been a puzzle 
since Greywall and Busch reported the specific heat data at the coverage 
$\rho = 0.178$ \AA$^{-2}$.\cite{godfrin,greywall89} At this coverage 
the adsorbed $^3$He atoms  form two layers and each layer solidifies
into a triangular lattice.
The first layer is a high-density solid and is considered not to contribute to 
the magnetism in the mK temperature region.
The  specific heat data down to 2 mK showed a prominent peak at 2.5 mK
as is expected from the conventional Heisenberg antiferromagnet (HAF) on 
the triangular lattice.
 But integration of their data extrapolated linearly to lower temperatures 
indicated that a half of the total magnetic entropy will  not be released 
down to the absolute zero temperature. 
Elser proposed that the second layer forms a $\sqrt{7} \times \sqrt{7}$ 
structure in registry with the first layer triangular lattice.\cite{elser89} 
 Then it decomposes into inequivalent A and B sites which may be magnetically 
decoupled.
Since  A sites form a kagom\'e lattice,  
 he  proposed that the missing entropy may be explained by the HAF on the 
kagom\'e lattice rather than that on 
the triangular lattice.
Inspired by this proposal many investigations were accomplished on 
the HAF on the kagom\'e lattice in recent years.\cite{kagome}
According to numerical studies of finite clusters the kagom\'e HAF shows 
a second peak in the specific heat at a low temperature, 
which might compensate the missing entropy.\cite{elser89,nakamura-m95}
On the other hand, Roger showed for a cluster with 16 spins that a second peak
 can be also produced  by  
effects of the multi-spin interactions which should be of considerable amount 
in solid $^3$He systems.\cite{roger90} 

Recent specific heat data down to 0.1mK 
at the coverage $\rho = 0.180$ \AA$^{-2}$, however, 
do not exhibit any sharp second peak in the temperature region 
where the kagom\'e HAF predicts.\cite{fukuyama96}
 Furthermore 
the estimate of the entropy released at temperatures above 0.1mK  
nearly exhausts the total 
magnetic entropy of the second layer. The peak at $T\simeq 2.5$ mK found 
by Greywall and 
Busch turned out to have a much broader tail in the low temperature region 
than was first thought. It is clear that these features of recently obtained
specific heat data cannot be explained in terms of either the HAF on the 
triangular lattice with nearest neighbor exchange couplings or 
that on the  kagom\'e lattice. 
So the behavior of the specific heat
remains as a puzzle though there is not missing entropy any longer.

It is well known that magnetic interactions due to ring exchanges 
involving more than two spins are important in solid $^3$He.
The famous {\it uudd} state in bcc solid $^3$He is incorporated by 
the four-spin exchange interaction.\cite{roger-hd83} 
The exchange frequencies in bcc solid $^3$He computed 
by using the path-integral Monte Carlo method  agree fairly well
with estimates from experiments.\cite{ceperley-j87}
For a monolayer of $^3$He at the coverage $\rho = 0.0785$ \AA$^{-2}$ 
on grafoil the cyclic-exchange frequencies were computed 
by Bernu et al.\ in a similar way.\cite{bernu-cl92}
  Their result shows that the three-spin exchange term
predominates the others at this coverage and the high-temperature specific
 heat is governed by 
the  two-, three- and  four-spin exchange interactions
while  the six-spin exchange term  as well 
is necessary to be taken into account to estimate 
the Weiss temperature properly.

The recent experimental results indicate that we have to take account of 
all spins in the second layer in order to understand its magnetism. 
We assume that the difference between A and B sites causes negligible effects 
on the translational symmetry of the magnetic 
interactions and that the magnetism of the second layer may be described by
 an $S=1/2$ spin system on the regular triangular lattice.
We employ the Hamiltonian with the two-, three- and four-spin 
exchange interactions as an effective Hamiltonian for the second layer
of adsorbed  $^3$He, though no estimates of exchange frequencies were reported 
for the second layer so far. 
This model may describe the $^3$He monolayer adsorbed on plated graphite 
as well.\cite{lusher-sc91,siqueira-lcs93}
Since three-spin exchange terms map into usual two-spin exchange couplings, 
the model is expressed  by the two-parameter Hamiltonian
\begin{eqnarray}
 H = J\sum_{<i,j>}^{\rm n.n.}
\mbox{\boldmath $\sigma$}_{i}\cdot\mbox{\boldmath $\sigma$}_{j}
    + K\sum_{p}h_{p}, 
\label{eq1}
\end{eqnarray}
where \mbox{\boldmath $\sigma_{i}$}/2 is the nuclear spin of the $^3$He atom 
at the site $i$ and 
$\sum_{p}$ denotes the sum over all possible diamonds composed of 
two nearest neighbor unit triangles. The four-spin exchange term 
of the diamond $p$ depicted in Fig.~1(a) reads
\begin{eqnarray}
 h_{p} = \sum_{< \alpha, \beta >}
       \mbox{\boldmath $\sigma$}_{\alpha}\cdot\mbox{\boldmath $\sigma$}_{\beta}
       + ( \mbox{\boldmath $\sigma$}_{a}\cdot\mbox{\boldmath $\sigma$}_{b} )
         ( \mbox{\boldmath $\sigma$}_{c}\cdot\mbox{\boldmath $\sigma$}_{d} )
                                                     \nonumber \\
       + ( \mbox{\boldmath $\sigma$}_{a}\cdot\mbox{\boldmath $\sigma$}_{d} )
         ( \mbox{\boldmath $\sigma$}_{b}\cdot\mbox{\boldmath $\sigma$}_{c} )  
       - ( \mbox{\boldmath $\sigma$}_{a}\cdot\mbox{\boldmath $\sigma$}_{c} )
         ( \mbox{\boldmath $\sigma$}_{b}\cdot\mbox{\boldmath $\sigma$}_{d} ),
\label{eq2}
\end{eqnarray}
where  $\sum_{< \alpha, \beta >}$ denotes the sum over all possible pairs 
of four spins on the diamond.  
The parameter $J$ and $K$ are related to the exchange frequency $J_n$ 
of the $n$-body cycle of the nearest neighbors as
\begin{eqnarray}
 J = -J_{2}/2 + J_{3}
\label{eq3}
\end{eqnarray}
and
\begin{eqnarray}
 K = -J_{4}/4.
\label{eq4}
\end{eqnarray}
It should be noted that the exchange frequency $J_n$ is always negative 
for any $n$.\cite{thouless65}
The magnitude of the two-body exchange frequency $J_{2}$ is expected 
to be large at a low enough 
atomic density of the layer but it decreases with density more rapidly 
than $J_{3}$.\cite{roger-hd83,roger90} 
Thus the parameter $J$ is expected to be positive at a low density 
while it should be 
 negative in the high density region. On the other hand $K$ is always positive.
We expect that a wide parameter region of the ratio of $J$ to $K$ may be
 realized by varying the coverage of $^3$He layers in a realistic system.
It is therefore useful to investigate what kind of the ground state occurs
in the system described by the Hamiltonian (\ref{eq1}). 
Calculations of the specific heat and the susceptibility of a 16-spin cluster
 were reported by Bernu et al.\ previously.\cite{bernu-cl92} 
\section{The ground state in the classical limit}
As a first step to study the spin system described by 
the Hamiltonian (\ref{eq1}), we examine the ground state 
in the classical limit. Since we neglect quantum fluctuations, 
we treat \mbox{\boldmath $\sigma_{i}$} 
 as a unit classical vector in the following. When $K=0$ the ground state 
of the system is well known,  that is,  the ferromagnetic state for $J<0$  
and the state with the so-called 120$^\circ$ structure for $J>0$.
On the other hand we cannot guess the ground state of the 
system intuitively when $J = 0$, 
since the Hamiltonian is governed by the unusual four-spin term. 
So we use a variational argument. We divide the total Hamiltonian into
a  sum of those of interpenetrating hexagons with 7 spins. That is
\begin{eqnarray}
 H =  {K\over 2}\sum_{h}H_{h},
\label{eq5}
\end{eqnarray}
where $H_{h}$ is the Hamiltonian of the hexagon  depicted in Fig.~1(b) and 
is expressed as a sum of Hamiltonians (\ref{eq2}) for six diamonds contained 
in the hexagon. Then we searched  numerically for the spin configuration 
which minimizes $H_{h}$. 
 It has turned out that the lowest-energy spin configuration is 
a four-sublattice spin
structure where any two spins on different sublattices 
make the relative angle $\alpha$ where $\cos \alpha = -1/3$ (see Fig.~2(b)).
We can divide the whole lattice into 
four sublattices such that four vertices of every diamond belong 
to different sublattices.
The spin configuration which minimizes a hexagon, therefore,
 can be extended to the whole lattice in such a way that the state 
has the lowest energy for all hexagons in eq.~(\ref{eq5}) consistently 
and so realizes the ground state 
of the whole lattice. The spin configuration has the four-sublattice structure,
where the spin vectors point the vertices of a tetrahedron if
we put their bottoms at its center, as shown in Fig.~2(b).
Hence we call this spin structure as "tetrahedral" structure. 
In fact we can prove in the same way that the tetrahedral structure is 
the ground state at least in the parameter region $-K/2 \le  J \le  2 K $.

For general values of $J$ and $K$ we determine the ground state 
in the mean-field 
approximation considering both three- and  four-sublattice spin structures. 
In a three-sublattice structure with the sublattices $a$, $b$, and $c$, 
the energy is expressed as 
\begin{eqnarray}
 E/N &=& (J+4K)(\mbox{\boldmath $\sigma$}_{a}\cdot\mbox{\boldmath $\sigma$}_{b}
         +  \mbox{\boldmath $\sigma$}_{b}\cdot\mbox{\boldmath $\sigma$}_{c} 
         +  \mbox{\boldmath $\sigma$}_{c}\cdot\mbox{\boldmath $\sigma$}_{a} )
                                                     \nonumber \\
        &+& 2K\{( \mbox{\boldmath $\sigma$}_{a}\cdot
                  \mbox{\boldmath $\sigma$}_{b} )
           ( \mbox{\boldmath $\sigma$}_{c}\cdot\mbox{\boldmath $\sigma$}_{a} ) 
         + ( \mbox{\boldmath $\sigma$}_{b}\cdot\mbox{\boldmath $\sigma$}_{c} )
           ( \mbox{\boldmath $\sigma$}_{a}\cdot\mbox{\boldmath $\sigma$}_{b} )
                                                      \nonumber \\
        &+& ( \mbox{\boldmath $\sigma$}_{c}\cdot\mbox{\boldmath $\sigma$}_{a} )
        ( \mbox{\boldmath $\sigma$}_{b}\cdot\mbox{\boldmath $\sigma$}_{c} ) \}
        + 3K.    
\label{eq6}
\end{eqnarray}
 In a four-sublattice structure with  the sublattices $a$, $b$, $c$ and $d$, 
we have
\begin{eqnarray}
 E/N &=& {1\over 2}(J+6K)\sum_{\alpha < \beta}
     \mbox{\boldmath $\sigma$}_{\alpha}\cdot\mbox{\boldmath $\sigma$}_{\beta}
                                                           \nonumber \\
     &+& K\{( \mbox{\boldmath $\sigma$}_{a}\cdot\mbox{\boldmath $\sigma$}_{b} )
         ( \mbox{\boldmath $\sigma$}_{c}\cdot\mbox{\boldmath $\sigma$}_{d} )  
      + ( \mbox{\boldmath $\sigma$}_{a}\cdot\mbox{\boldmath $\sigma$}_{c} )
        ( \mbox{\boldmath $\sigma$}_{b}\cdot\mbox{\boldmath $\sigma$}_{d} )
                                                      \nonumber \\
     &+& ( \mbox{\boldmath $\sigma$}_{a}\cdot\mbox{\boldmath $\sigma$}_{d} )
        ( \mbox{\boldmath $\sigma$}_{b}\cdot\mbox{\boldmath $\sigma$}_{c} ) \}.
\label{eq7}
\end{eqnarray}
  
In above equations $ \mbox{\boldmath $\sigma$}_{a}$ denotes
the spin vectors on the sublattice $a$ etc.
We determined the ground state spin configuration by comparing 
the energies of the two configurations which minimize 
eqs.~(\ref{eq6}) and (\ref{eq7}), respectively.
Minimization of  eqs.~(\ref{eq6}) and  (\ref{eq7}) were done numerically. 
As a result we obtained four ground state phases with following 
spin structures:
\begin{enumerate} 
\item
the 120$^\circ$ structure with $E/N=-3(J+K)/2$ for $0<K<3J/25$,
\item
the tetrahedral structure with $E/N=-J-17K/3$ for $0<3J/25<K$  
     and  $0<-3J/8<K$,
\item
the {\it uuud} structure with $E/N=-3K$ for $0<-J/8<K<-3K/8$,
\item
the perfect ferromagnetism with $E/N=3J+21K$ for $0<K<-J/8$.
\end{enumerate}
 The {\it uuud} structure has four sublattices and spins on 
the three sublattices are aligned parallel while spins on the other one 
 is antiparallel to them. The ground-state spin structures are
 depicted 
in Fig.~2.

The most interesting and novel one among the obtained ground-state 
spin structures is the tetrahedral structure.
It is remarkable that this structure has a scalar chiral long-range order.
 The scalar chirality in the unit cell may be defined as 
\begin{eqnarray}
 \chi_{\rm S} &=& 
          \mbox{\boldmath $\sigma$}_{a}\cdot
         ( \mbox{\boldmath $\sigma$}_{b}\times \mbox{\boldmath $\sigma$}_{c} )
             + \mbox{\boldmath $\sigma$}_{b}\cdot
         ( \mbox{\boldmath $\sigma$}_{d}\times \mbox{\boldmath $\sigma$}_{c} )
                                           \nonumber \\
     &+&   \mbox{\boldmath $\sigma$}_{c}\cdot
         ( \mbox{\boldmath $\sigma$}_{d}\times \mbox{\boldmath $\sigma$}_{a} )
             + \mbox{\boldmath $\sigma$}_{d}\cdot
         ( \mbox{\boldmath $\sigma$}_{b}\times \mbox{\boldmath $\sigma$}_{a} ),
\label{eq8}
\end{eqnarray}
which is an order parameter of Ising-type with a discrete $Z_2$ symmetry. 
Since the tetrahedral ground state has a finite value of $\chi_{\rm S}$, 
$16\sqrt{3}/9$, 
the system may undergo a phase transition of Ising universality 
at a finite temperature in spite of the isotropic 
rotational symmetry inherent in the Hamiltonian.
 This contrasts with the 120$^\circ$ structure which has a vector 
chiral order and may cause only topological phase transitions at finite 
temperatures.\cite{kawamura-m84}

\section{Phase diagram in  the magnetic field}

Application of a magnetic field to frustrated spin systems is known to cause 
interesting magnetization processes and/or various phase transitions.
 First we consider the case with $K=0$ and $J>0$, i.e. the HAF on the  
triangular lattice with nearest neighbor coupling $J$. It is well known that 
 two phase transitions occur under the magnetic field 
at T=0.\cite{kawamura-m85}
When the magnitude of the field $H$ is less than $3J$ (the magnetic moment
of a spin is assumed as unity), The ground state  is  the so-called umbrella 
state with a coplanar spin 
configuration where the spins on a sublattice orient 
downwards (the magnetic field
 is assumed to be applied upwards)
 and those on the other two
sublattices orient upwards but canted to the field by an angle (see Fig.~3(a)).
The angle decreases with the magnetic field and vanishes at $H=3J$.
For  $3J<H<9J$, spins on all sublattices are canted to the field but 
two of them are parallel to each other (see Fig.~3(c)).
We call the phase with this spin structure as c$_3$ phase.
 Spins are perfectly polarized for $H>9J$.
At $H=3J$ the spins are collinear, i.e. those on two sublattices 
are oriented upwards
 while those on the other sublattice point downwards (see Fig.~3(b)). 
We call this spin structure as {\it uud} structure.
 It was shown that the phase with the {\it uud} structure expands and is 
realized in a finite 
range of the magnetic field aided by the thermal 
or the quantum mechanical fluctuations.\cite{kawamura-m85,chubukov-g91}
   
We searched for the ground state for general values of $J$ and $K$ numerically 
minimizing eqs.~(\ref{eq6}) and (\ref{eq7}) added by
 $ - ( \sigma_{a}^{z}+\sigma_{b}^{z}+\sigma_{c}^{z})H/3$
and 
$ - ( \sigma_{a}^{z}+\sigma_{b}^{z}+\sigma_{c}^{z}+\sigma_{d}^{z})H/4$, 
respectively.
We found seven ground-state phases in total. They are the following:
\begin {enumerate}
\item 
the umbrella phase, which appears for a weak magnetic field
 when $J>0$ and $K$ is small.
\item
the {\it uud} phase which extends around $H\simeq 3J(>0)$ and for small $K$.
\item
the  c$_3$ phase, 
 which appears for a wide range of the parameters both
for positive and negative $J$ near the boundary of the fully polarized phase.
\item 
the weak field four-sublattice phase, the spin structure is an analogue
of the umbrella structure but is not coplanar. The spins on a 
sublattice point downwards and those on the other three
sublattices orient upwards but canted to the field by an angle 
(see Fig.~3(d)). 
There is a  three-fold rotational symmetry about the magnetic field axis. 
We call this structure "mushroom" structure. This phase appears 
for large $K$ and small $H$.
\item
the {\it uuud} phase which appears for small $|J|$  and rather weak field.
\item 
the high field four-sublattice phase with the structure analogous to 
the c$_3$ phase.
Spins on three sublattices are parallel to each other and all spins are canted 
to the field (see Fig.~3(e)).
We call this phase as c$_4$ phase.
\item
the ferromagnetic phase where the spins are perfectly polarized.
 This phase appears for $H \ge 9(J+8K)$.

\end{enumerate}
Above spin structures are shown in Fig.~3. 
The phase diagram is depicted in Fig.~4 for  $J>0$  and in Fig.~5 for  $J<0$.
We adopted a scale as $J+8K=1$ for $J>0$ and $-2J+8K=1$ for $J<0$ 
for convenience.
The full lines denote the first order transitions while dotted lines 
the second order ones.
The transitions between the three- and the four-sublattice structures 
are necessarily of the first order.
 The transitions between the structures with the same sublattice 
structures are of the second order in most cases. An exception is the boundary
between the mushroom and the {\it uuud} phases for $J<0$, where the 
 phase change occurs with a magnetization jump.
The critical field $H_{\rm C}$ at the
second order transition points are determined analytically and
 tabulated in table 1.

There is a tetracritical point at $H_{\rm C}=6J=24K$ when $J=4K$.
 Magnetization process may show
peculiar behaviors around this point.
As an example magnetization per spin $m$ is shown in Fig.~6 as a function of 
the applied field for $J=0.4$ and $K=0.75$. 
In this case the system undergoes six phase transitions with variation
 of $H$. There appear two plateaus in the magnetization curve at $m=1/3$
 and $m=1/2$. The 1/3- and 1/2-plateaus correspond to the {\it uud} and 
{\it uuud} phases, respectively. Furthermore the curve shows magnetization 
jumps at the first-order phase-transition points. Another magnetization 
curve is shown in Fig.~7 for the parameter $J=-0.1$ and $K=0.1$.

\section{Summary and discussion}

We have shown that various ground-state phases occur 
on the triangular lattice 
by coexistence of the two- and the four-spin exchange interactions. 
We have found new phases with four sublattices, i.e. the tetrahedral structure 
and the {\it uuud} structure. 
Since we have assumed that the ground states have either the three- or the 
four-sublattice structure, there remains a possibility that a state with
a longer period may be more stable than those obtained above. 
In some cases, however, we are certain that the present
 results are rigorous in the classical limit.
That is, the case when $K=0$ (the pure Heisenberg model), 
and the case when $J=0$ and $H=0$. So it is 
quite reasonable that the obtained phases are stable in the regions close 
to these limiting cases. In fact we could prove that the tetrahedral phase 
is stable for small $|J|$. 

It is remarkable
that the phase with scalar chirality appears when the four-spin exchange 
dominates.
This suggests the existence of a phase transition of the Ising universality 
at a finite temperature. 
Our preliminary Monte Carlo simulations in the classical limit 
clearly exhibit divergence of the specific heat at $T \simeq K$ for $J=0$.
The present model seems to be the first example of a realistic  spin system
in two dimensions
with full rotational symmetry which undergoes a chiral phase transition.
Detailed analysis of the finite-temperature properties will be reported
 elsewhere.\cite{mmoi-k97} 

In the region where the competition between 
the two- and the four-spin interactions are serious, more study is necessary 
to be
 conclusive. In fact we found in Monte Carlo simulations that various states 
with long periods and zero magnetization have quite close energy to that of 
the {\it uuud} structure. Furthermore, it can be shown that a new $120^\circ$ 
structure state with nine sublattices has the same energy as the {\it uuud} 
structure. This indicates that the frustration of the system is 
strong in the phase with {\it uuud} structure due to the competition 
between the two- and the four-spin interactions. 
The application of the magnetic field would select the 
{\it uuud} structure from the degenerate ground states, since it has the 
largest magnetization among them. We hence believe that the ground-state 
phase diagram under the magnetic field (Fig.~5) will not be changed seriously 
by the appearance of long-period states. 

Effects of quantum fluctuations, which we have completely neglected 
in the present study,  are surely 
 very important since the magnitude of the nuclear spin is 1/2 for $^3$He. 
It is still not
quite established that even the well-known 120$^\circ$ structure is stable 
against quantum fluctuations when $K=0$ and $J>0$ (HAF), 
though recent numerical studies seem to 
support for the stability.\cite{triangle} It is possible 
that the coexistence of the two- and the four-spin interactions destabilizes 
the ordered ground state and helps the quantum fluctuations to destroy it.
The stability of the ground states against the zero-point spin wave 
fluctuations are presently under investigation and will be reported 
elsewhere.\cite{niki-k97} 

Finally we compare our results with the existing experimental ones,
though it is still not quite clear whether the model is relevant 
to describe  $^3$He layers. The specific heat 
data at low temperatures exist only at two different
values of the coverage, i.e. $\rho = 0.180$ \AA$^{-2}$ and 
$0.228$ \AA$^{-2}$.\cite{fukuyama96,Morishita96} 
No sharp peak  was observed at both coverages.  
The system with   $\rho = 0.228$ \AA$^{-2}$ 
clearly corresponds to a negative $J$ since it shows about 60 percent
of full polarization under a very small magnetic field.\cite{Schiffer93} 
On the other hand the parameter $J$ of the system with $\rho=0.180$ \AA$^{-2}$ 
is still unclear. At this coverage the system has been thought to be 
antiferromagnetic since it has a negative Weiss temperature $\theta$. 
However a negative $\theta$ does not imply a positive $J$ since 
$\theta = -6( J + 6K )$ holds in the present model.
 If the system corresponds to a positive $J$, the ground state at this 
coverage 
should be  the positive large-$J$ phase or a new quantum phase caused by 
competitions between the two- and the 
four-spin intearctions and quantum fluctuations, since the system does not show
 a finite-temperature phase transition. 
If this is the case, there may occur the phase with a
chiral long-range order between two coverages $\rho=0.180$ \AA$^{-2}$ and 
$0.228$ \AA$^{-2}$ (if the density of the second layer varies continuously), 
and we may observe a sharp peak in the specific heat 
by varying the coverage of the adsorbed layers.
Another possibility is that the system with $\rho=0.180$ \AA$^{-2}$ 
has a negative $J$ whose value corresponds to the {\it uuud} ground state 
in the mean field theory. In this parameter region we have seen that
the strong frustration leads to the existence of various states 
with long periods whose energies are quite 
close to that of the ground state. Then the effects of quantum fluctuations
  may realize a novel ground state without magnetic long-rage order. 
 Observations under a magnetic field will be useful to determine 
the phase that the system belongs to. 
Above discussion is based on the assumption that 
 more-than-four-spin exchange interactions are not important. If this is 
not the case  it is necessary to investigate a spin model
with more-than-four-spin exchange interactions in search for possible ground 
states. 

\section{acknowledgement}
The authors greatly appreciate Hiroshi Fukuyama for guiding them to 
the present problem and for  stimulating discussions and useful advice.
Thanks are also due to K. Niki, and H. Sakamoto for useful discussions.

\newpage
\centerline{Table}

\begin{table}[b]
\begin{center}
\begin{tabular}{lcc}
 Phase1 & Phase2 & $H_{\rm C}$  \\
\hline
  umbrella       &   {\it uud}    &  $3(J-4K)$  \\
  {\it uud}      &   c$_3$  &  $3(J+4K)$  \\
  ferromagnetic  &   c$_3$  &  $9(J+8K)$  \\
  mushroom       &   {\it uuud}   &  $4(J+2K)$  \\
  {\it uuud}     &   c$_4$  &  $4(J+8K)$  \\
  \end{tabular}

\caption{$H_{\rm C}$ at the second order transition point.}
\end{center}
\end{table}

\centerline{Figure Captions}
\begin{figure}
 \caption{(a) A unit diamond described by the four-spin exchange 
  Hamiltonian (2).
  (b) A unit hexagon described by $H_{h}$ in eq.~(5).
   \label{fig:fig1}
         }
\end  {figure}
\begin{figure}
  \caption{The ground state spin structures: 
(a) the 120$^\circ$, (b) the tetrahedral  
 and (c) the {\it uuud} structure. The ferromagnetic spin structure 
 is an obvious one. 
  \label{fig:fig2}
          }
\end  {figure}
\begin{figure}
 \caption{The ground state spin structures in the magnetic field:
   (a) the umbrella, (b) the {\it uud},
   (c) the c$_{3}$,  (d) the mushroom  and 
   (e) the c$_{4}$ structure. The {\it uuud} structure is shown in Fig.~2.
  The magnetic field is assumed to be applied upwards. 
 \label{fig:fig3}
         }
\end  {figure}
\begin{figure}
 \caption{The ground state phase diagram for $J>0$. The parameters $J$
  and $K$ are scaled as $J+8K=1$. The phases are  labelled by 
 u(umbrella), m(mushroom)  and 
f(perfect ferromagnetism). Other labels are self-evident.
The full and the dotted lines indicate the phase boundaries of the first 
and the second orders, respectively.
  \label{fig:fig4} 
         }
\end  {figure}
\begin{figure}
 \caption{The ground state phase diagram for $J<0$.
      The parameters $J$ and $K$ are scaled as $-2J+8K=1$. 
The labels and the lines are the same as in Fig.~4. 
 \label{fig:fig5}
         }
\end  {figure}
\begin{figure}
 \caption{The magnetization curve for $J=0.4$ and $K=0.75$. 
The system undergoes three first-order transitions and three second-order 
ones. The c$_3$ phase appears twice.
         }
\end  {figure}
\begin{figure}
 \caption{The magnetization curve for $J=-0.1$ and $K=0.1$. 
The system undergoes two first-order transitions and two second-order 
ones. 
         }
\end  {figure}
\end{document}